\documentclass[12pt]{article}
\usepackage{epsfig}
\usepackage{amsmath}
\usepackage{hhline}
\usepackage{amssymb}
\usepackage{cite}
\usepackage{times}
\usepackage{color}
\usepackage{lineno}

\newlength{\dinwidth}
\newlength{\dinmargin}
\begin{document}

\pagestyle{empty}
\begin{titlepage}

\noindent

\begin{flushleft}
DESY 05-093 \hfill ISSN 0418-9833 \\
July 2005
\end{flushleft}

\vspace{2cm}

\vspace{2cm}

\begin{center}
  \Large
  {\bf\boldmath 
A Determination of Electroweak Parameters at HERA}

  \vspace*{1cm}
    {\Large H1 Collaboration} 
\end{center}

\begin{abstract}
\noindent
Using the deep inelastic  $e^+p$ and $e^-p$ charged and neutral current scattering  cross sections
previously published, a combined electroweak and QCD
analysis is performed to determine electroweak parameters accounting
for their correlation with parton distributions.
The data used have been collected by the H1 experiment in $1994$-$2000$
and correspond to an integrated luminosity of $117.2\,{\rm pb}^{-1}$. 
A measurement is obtained of the $W$ propagator mass
in charged current $ep$ scattering. The weak mixing angle $\sin^2\!\theta_W$
is determined in the on-mass-shell
renormalisation scheme. 
A first measurement at HERA is made of the light quark weak couplings to
the $Z^0$ boson and
a possible contribution of right-handed isospin components to the weak
couplings is investigated. 

\noindent
\end{abstract}

\vspace{5mm}
\begin{center}
(To be submitted to Phys.\ Lett.\ B.)
\end{center}

\end{titlepage}

\begin{flushleft}

A.~Aktas$^{10}$,               
V.~Andreev$^{26}$,             
T.~Anthonis$^{4}$,             
S.~Aplin$^{10}$,               
A.~Asmone$^{34}$,              
A.~Astvatsatourov$^{4}$,       
A.~Babaev$^{25}$,              
S.~Backovic$^{31}$,            
J.~B\"ahr$^{39}$,              
A.~Baghdasaryan$^{38}$
% ...

\bigskip{\it
% ...
$ ^{4}$ Inter-University Institute for High Energies ULB-VUB, Brussels;
          Universiteit Antwerpen, Antwerpen; Belgium$^{ c}$ \\
% ...
$ ^{10}$ DESY, Hamburg, Germany \\
% ...
$ ^{25}$ Institute for Theoretical and Experimental Physics,
          Moscow, Russia$^{ l}$ \\
 $ ^{26}$ Lebedev Physical Institute, Moscow, Russia$^{ e}$ \\
% ...
$ ^{31}$ Faculty of Science, University of Montenegro,
          Podgorica, Serbia and Montenegro$^{ e}$ \\
% ...
$ ^{34}$ Dipartimento di Fisica Universit\`a di Roma Tre
          and INFN Roma~3, Roma, Italy \\
% ...
$ ^{38}$ Yerevan Physics Institute, Yerevan, Armenia \\
 $ ^{39}$ DESY, Zeuthen, Germany \\
% ...
$ ^c$ Supported by FNRS-FWO-Vlaanderen, IISN-IIKW and IWT
      and  by Interuniversity
Attraction Poles Programme,
      Belgian Science Policy \\
% ...
$ ^e$ Supported by the Deutsche Forschungsgemeinschaft \\
% ...
$ ^l$ Partially Supported by Russian Foundation
      for Basic Research, grant    no. 00-15-96584 \\
}

\end{flushleft}

\newpage
\pagestyle{plain}

\section{Introduction}
The deep inelastic scattering (DIS) of leptons off nucleons has played 
an important role in revealing the structure of matter, in the discovery
of weak neutral current interactions and 
in the foundation of the Standard Model (SM) as the theory of strong and 
electroweak (EW) interactions.
At HERA, the first lepton-proton collider ever built, the study of DIS 
has been pursued since 1992 over a wide kinematic range. 
In terms of $Q^2$, the negative four-momentum transfer squared,
the kinematic coverage includes the region
where the electromagnetic and weak interactions become
of comparable strength.
Both charged current (CC) and neutral current (NC) interactions 
occur in $ep$ collisions and are studied by the two collider 
experiments H1 and ZEUS. Many QCD analyses of HERA data have been performed
to determine the strong interaction coupling constant 
$\alpha_s$~\cite{h1alphas,zeusalphas,zeusalphasnew}
and parton distribution functions (PDFs)~\cite{zeusalphas,h1ep300,h1ep320}.
In EW analyses, the $W$ boson mass value has been determined from the charged 
current data at high $Q^2$~\cite{h1ep300,h1cc94,h1nccc96,h1em320,zeusccep300,
zeusccem320,zeusccep320}. Previously the QCD and EW sectors were analysed independently.

Based solely on the precise data recently published by 
H1~\cite{h1alphas,h1ep300,h1em320,h1ep320}, a combined QCD and EW 
analysis is performed here for the first time and parameters of the
electroweak theory are determined.
The data have been taken by the H1 experiment in the first phase of operation of 
HERA (HERA-I) with unpolarised $e^+$ and $e^-$ beams and correspond to 
an integrated luminosity of $100.8\,{\rm pb}^{-1}$ for $e^+p$ 
and $16.4\,{\rm pb}^{-1}$ for $e^-p$ respectively.
A measurement is made of the $W$ mass in the space-like region from 
the propagator mass ($M_{\rm prop}$) in charged current scattering.
The masses of the $W$ boson ($M_W$) and top quark ($m_t$) and 
the weak mixing angle ($\sin^2\!\theta_W$) are determined 
within the electroweak ${\rm SU}(2)_L \times {\rm U}(1)_Y$ Standard Model.
The vector and axial-vector weak couplings of
the light ($u$ and $d$) quarks to the $Z^0$ boson are measured
for the first time at HERA.
These results are complementary to determinations of EW parameters at LEP,
the Tevatron and low energy experiments~\cite{pdg04}.

\section{Charged and Neutral Current Cross Sections}\label{sec:smnccc}

\subsection{Charged Current Cross Section}

The charged current interactions, 
$e^\pm p \rightarrow \overline{\nu}_e^{\mbox{\tiny
\hspace{-3mm}\raisebox{0.3mm}{(}\hspace{2.5mm}\raisebox{0.3mm}{)}}}X$,
are mediated by the exchange of a $W$ boson in the $t$ channel.  
The measured cross section for unpolarised beams after correction for QED
radiative effects~\cite{ccncrad1,ccrad2,heracles} can be expressed as
%
\begin{eqnarray}
 \frac{d^2\sigma^{\rm CC}(e^\pm p)}{dxdQ^2}&=&
\frac{G^2_F}{2\pi x}\left[\frac{M^2_W}{M^2_W+Q^2}\right]^2\phi^\pm_{CC}(x,Q^2)
\left(1+\Delta^{\pm, weak}_{CC}\right)\,,\label{eqn:xscc}\\
{\rm with} \hspace{5mm} \phi^\pm_{CC}(x,Q^2)&=&
\frac{1}{2}\left[Y_+W_2^\pm (x,Q^2)\mp Y_-xW_3^\pm (x,Q^2)-y^2W_L^\pm (x,Q^2)\right]\,.
\end{eqnarray}
Here $G_F$ is the Fermi constant accounting for radiative corrections to 
the $W$ propagator as measured in muon decays and $\Delta^{\pm,weak}_{CC}$ 
represents the other weak vertex and box corrections, which amount to 
a few per mil~\cite{wh} and are neglected.
The term $\phi^\pm_{CC}$~\cite{h1ep300}
contains the structure functions $W^\pm_2$, $xW^\pm_3$ and
$W^\pm_L$. The factors $Y_\pm$ are defined as $Y_\pm =  1\pm (1-y)^2$ and
$y$ is the inelasticity variable which is related to Bjorken $x$,
$Q^2$ and the centre-of-mass energy squared $s$ by $y=Q^2/xs$.

Within the SM, the CC cross section in Eqn.(\ref{eqn:xscc}) can
be expressed in the so-called on-mass-shell (OMS) scheme~\cite{oms}
replacing the Fermi constant $G_F$ with:
%
\begin{equation}
G_F=\frac{\pi\alpha}{\sqrt{2} M^2_W\left(1-\frac{M^2_W}{M^2_Z}\right)}\frac{1}{1-\Delta r}\,,
\label{eqn:gf}
\end{equation}
where $\alpha\equiv \alpha(Q^2=0)$ is the fine structure constant and 
$M_Z$ is the mass of the $Z^0$ boson.
The term $\Delta r$ contains one-loop and leading higher-order EW radiative 
corrections. The one-loop contributions can be expressed as~\cite{wh}
%
\begin{equation}
\Delta r=\Delta \alpha -\frac{\cos^2\!\theta_W}{\sin^2\!\theta_W}\Delta \rho
+\Delta r_{\rm rem}\label{eqn:dr}\,.
\end{equation}
The first term $\Delta \alpha$ is the fermionic part of 
the photon vacuum polarisation. It has a calculable leptonic contribution 
and an uncalculable hadronic component
which can however be estimated using $e^+e^-$ data~\cite{dalpha}. 
Numerically these two contributions are of similar size and
have a total value of $0.059$~\cite{eprc} when evaluated at $M^2_Z$. 
The quantity $\Delta \rho$ 
arises from the large mass difference between the top and bottom quarks 
in the vector boson self-energy loop:
%
\begin{equation}
\Delta \rho=\frac{3\alpha}{16\pi \sin^2\!\theta_W\cos^2\!\theta_W}\frac{m^2_t}{M^2_Z}\,,\label{eqn:drho}
\end{equation}
after neglecting the mass of the bottom quark.
The second term in Eqn.(\ref{eqn:dr}) has a numerical value of about $0.03$. 
The last term $\Delta r_{\rm rem}$ is numerically smaller ($\sim 0.01$). 
It contains the remaining contributions including those with logarithmic 
dependence on $m_t$ and the Higgs boson mass $M_H$. Leading higher-order terms
proportional to $G_F^2 m_t^4$ and $\alpha\alpha_s$ are included as well.
In Eqns.(\ref{eqn:dr},\ref{eqn:drho}) and the OMS scheme, it is understood 
that 
%
\begin{equation}
\sin^2\!\theta_W=1-M^2_W/M^2_Z.
\label{eqn:s2theta}
\end{equation}

In the quark parton model (QPM), 
the structure functions $W^\pm_2$ and $xW^\pm_3$ may be interpreted as
lepton charge dependent sums and differences of quark and anti-quark 
distributions and are given by
%
\begin{equation}
W^+_2=x(\overline{U}+D),\hspace{3mm} xW^+_3=x(D-\overline{U}),\hspace{3mm}
W^-_2=x(U+\overline{D}),\hspace{3mm} xW^-_3=x(U-\overline{D})\,,
\end{equation}
whereas $W^\pm_L=0$. 
The terms $xU$, $xD$, $x\overline{U}$ and $x\overline{D}$
are defined as the sum of up-type, of down-type and of their anti-quark-type 
distributions, i.e. below the $b$ quark mass threshold:
%
\begin{equation}
xU=x(u+c) , \hspace{3mm}
xD=x(d+s),\hspace{3mm}
x\overline{U}= x(\overline{u}+\overline{c}),\hspace{3mm}
x\overline{D}= x(\overline{d}+\overline{s})\,.
\end{equation}
In next-to-leading-order (NLO) QCD and the $\overline{{\rm MS}}$ 
renormalisation scheme~\cite{msbar},
these simple relations do not hold any longer
and $W^\pm_L$ becomes non-zero.
Nevertheless the capability of the CC cross sections to probe up- and 
down-type quarks remains.

\subsection{Neutral Current Cross Section}

The NC interactions, $e^\pm p\rightarrow e^\pm X$, are mediated by
photon ($\gamma$) or $Z^0$ exchange in the $t$ channel.
The measured NC cross section with unpolarised beams 
after correction for QED radiative effects~\cite{ccncrad1,ncrad2,heracles} 
is given by
%
\begin{eqnarray}
\frac{d^2\sigma^{\rm NC}(e^\pm p)}{dxdQ^2}&=&
\frac{2\pi\alpha^2}{xQ^4}\phi^\pm_{NC}(x,Q^2)
\left(1+\Delta^{\pm,weak}_{NC}\right)\,,\label{eqn:xsnc}\\
{\rm with}\hspace{5mm} \phi^\pm_{NC}(x,Q^2)&=&
Y_+\tilde{F}_2(x,Q^2)\mp Y_-x\tilde{F}_3(x,Q^2)-y^2\tilde{F}_L(x,Q^2)\,,
\end{eqnarray}
where $\Delta^{\pm,weak}_{NC}$ represents weak radiative corrections 
which are typically less than $1\%$ and never more than $3\%$.
The NC structure function term $\phi^\pm_{NC}$~\cite{h1ep300} is
expressed in terms of the generalised structure functions 
$\tilde{F}_2$, $x\tilde{F}_3$ and $\tilde{F}_L$.
The first two can be further decomposed as~\cite{klein}
%
\begin{eqnarray}
 \label{f2p}
 \tilde{F}_2  \equiv & F_2 & - \ v_e  \ \frac{\kappa  Q^2}{(Q^2 + M_Z^2)}
  F^{\gamma Z}_2  \,\,\, + (v_e^2+a_e^2)  
 \left(\frac{\kappa  Q^2}{Q^2 + M_Z^2}\right)^2 F^Z_2\,, \\
 \label{f3p}
 x\tilde{F}_3    \equiv &      & - \ a_e  \ \frac{\kappa  Q^2}{(Q^2 + M_Z^2)} 
 xF^{\gamma Z}_3 + \,\, (2 v_e a_e) \,\,
 \left(\frac{\kappa  Q^2}{Q^2 + M_Z^2}\right)^2  xF^Z_3\,.
\end{eqnarray} 
Here 
%
\begin{equation}
\kappa^{-1}=\frac{2\sqrt{2}\pi\alpha}{G_FM^2_Z}\,,
\end{equation}
in the modified on-mass-shell (MOMS) scheme~\cite{moms}, in which
all EW parameters can be defined in terms of $\alpha$, $G_F$ and $M_Z$ 
(besides fermion masses and quark mixing angles), or
%
\begin{equation}
\kappa^{-1}
=4\frac{M_W^2}{M_Z^2}\left(1-\frac{M_W^2}{M_Z^2}\right)(1-\Delta r)
\label{eqn:kappa}
\end{equation}
in the OMS scheme.
The quantities $v_e$ and $a_e$ are the
vector and axial-vector weak couplings of the electron
to the $Z^{0}$~\cite{pdg04}.
In the bulk of the HERA phase space, $\tilde{F}_2$ is dominated by 
the electromagnetic structure function $F_2$
originating from photon exchange only. The functions $F^Z_2$ and $xF^Z_3$
are the contributions to $\tilde{F}_2$ and $x\tilde{F}_3$ from $Z^0$ 
exchange and the functions $F_2^{\gamma Z}$ and $xF_3^{\gamma Z}$ are 
the contributions from $\gamma Z$ interference. 
These contributions only become important at large values of $Q^2$.

In the QPM, the longitudinal structure function $\tilde{F}_L$ equals zero 
and the structure functions $F_2$,
$F_2^{\gamma Z}$ and $F_2^Z$ are related to the sum of the quark
and anti-quark momentum distributions, $xq$ and $x\overline{q}$, 
%
\begin{equation}
 \label{eq:f2}
 [F_2,F_2^{\gamma Z},F_2^{Z}] = x \sum_q 
 [e_q^2, 2 e_q v_q, v_q^2+a_q^2] 
 \{q+\overline{q}\}, 
\end{equation}
whereas the structure functions $xF_3^{\gamma Z}$ and $xF_3^Z$ 
are related to their
difference,
%
\begin{equation}
 \label{eq:xf3}
 [ x F_3^{\gamma Z},x F_3^{Z} ] = 2x \sum_q 
 [e_q a_q, v_q a_q]
 \{q -\overline{q} \}\,.
\end{equation}
In Eqns.(\ref{eq:f2},\ref{eq:xf3}) $e_q$ is the electric charge of 
quark $q$, and $v_q$ and $a_q$ are, respectively, the vector and
axial-vector weak coupling constants of the quarks to the $Z^0$:
%
\begin{eqnarray}
v_q&=&I^3_{q,L}-2e_q\sin^2\!\theta_W, \label{eqn:vf}\\
a_q&=&I^3_{q,L}\label{eqn:af}
\end{eqnarray}
where $I^3_{q,L}$
is the third component of the weak isospin.

The weak radiative corrections $\Delta^{\pm,weak}_{NC}$ in 
Eqn.(\ref{eqn:xsnc}) correspond effectively to modifications of 
the weak neutral current couplings to so-called dressed couplings
 by four weak form factors
$\rho_{eq}$, $\kappa_e$, $\kappa_q$ and $\kappa_{eq}$~\cite{wh}.
The form factor $\rho_{eq}$ has a numerical value very close to $1$
for $Q^2\lesssim 10\,000\,{\rm GeV}^2$ and only at very high $Q^2$ 
a deviation of a few percent is reached~\cite{wh}. 
The form factors $\kappa_{e,q,eq}$ fall strongly with $Q^2$~\cite{wh} and
approach unity where the $\gamma Z$ and $Z^0$ contributions become significant.
Given the current precision of the data used (Section~\ref{sec:datafit}), 
in the following analysis $\rho_{eq}=1$ is assumed and the weak mixing angle 
in Eqn.(\ref{eqn:vf}) is replaced by an effective one, 
$\sin^2\!\theta_W^{\rm eff}=\kappa_q(1-M_W^2/M_Z^2)$, where $\kappa_q$ is 
assumed to be flavour independent and equal to the universal 
part of the form factors~\cite{eprc}.

\section{Data Sets and Fit Strategies}\label{sec:datafit}
The analysis performed here uses (as in~\cite{h1ep320}) the following
H1 data sets:
two low $Q^2$ data sets ($1.5\leq Q^2\leq 150\,{\rm GeV}^2$)~\cite{h1alphas}, 
three high $Q^2$ NC data sets 
($100\leq Q^2\leq 30\,000\,{\rm GeV}^2$)~\cite{h1ep300,
h1ep320,h1em320} and three high $Q^2$ CC data
sets ($300\leq Q^2\leq 15\,000\,{\rm GeV}^2$)~\cite{h1ep300,h1ep320,h1em320}.
These data cover a Bjorken $x$ range from $3\cdot 10^{-5}$ to $0.65$ 
depending on $Q^2$.

The low $Q^2$ data are dominated by systematic uncertainties which have
a precision down to $2\%$ in most of the covered region. The high $Q^2$
data on the other hand are mostly limited by the statistical
precision which is up to $30\%$ or larger for 
$Q^2\gtrsim 10\,000\,{\rm GeV}^2$.

The combined EW-QCD analysis follows the same fit procedure as used 
in~\cite{h1ep320}.
The QCD analysis is performed using 
the DGLAP evolution equations~\cite{dglap} at NLO~\cite{furmanski} 
in the $\overline{{\rm MS}}$ renormalisation scheme.
All quarks are taken to be massless.

Fits are performed to the measured cross sections 
assuming the strong coupling constant to be equal to 
$\alpha_s(M_Z)=0.1185$.
The analysis uses an $x$-space program developed within
the H1 Collaboration~\cite{qcdfit}.
In the fit procedure, a $\chi^2$ function which is defined
in~\cite{h1alphas} is minimised. 
The minimisation takes into account correlations between
data points caused by systematic uncertainties~\cite{h1ep320}.

In the fits, five PDFs -- gluon, $xU$, $xD$, $x\overline{U}$ and 
$x\overline{D}$ -- are defined by 10 free parameters as in~\cite{h1ep320}. 
Table~\ref{tab:fits} shows an overview of various fits that are
performed in the present paper to determine different EW parameters.
For all fits, the PDFs obtained here are consistent with those from
the H1 PDF 2000 fit~\cite{h1ep320}.
For more details refer to~\cite{bpthesis}.

\begin{table*}[htb]
  \renewcommand{\doublerulesep}{0.4pt}
  \renewcommand{\arraystretch}{1.2}
 \vspace{-0.1cm}

\begin{center}
    \begin{tabular}{|l|c|c|}
      \hline
 Fit  & \multicolumn{2}{|c|}{Fixed parameters} \\ \cline{2-3}
           & CC & NC \\\hline
$G$-$M_{\rm prop}$-${\rm PDF}$  & $-$ & 
  $\alpha, G_F, M_Z$ \\
$M_{\rm prop}$-${\rm PDF}$  & 
  $G_F$ & $\alpha, G_F, M_Z$ \\ \hline
$M_W$-${\rm PDF}$  & 
  \multicolumn{2}{|c|}{$\alpha, M_Z, m_t, M_H$} \\
$m_t$-${\rm PDF}$  & 
  \multicolumn{2}{|c|}{$\alpha, M_Z, M_W, M_H$} \\\hline
$v_u$-$a_u$-$v_d$-$a_d$-${\rm PDF}$  &
 $G_F, M_W$ & $\alpha, M_Z, M_W$ \\
$v_u$-$a_u$-${\rm PDF}$  &
 $G_F, M_W$ & $\alpha, M_Z, M_W, v_d, a_d$ \\
$v_d$-$a_d$-${\rm PDF}$  &
 $G_F, M_W$ & $\alpha, M_Z, M_W, v_u, a_u$ \\
$I^3_{u,R}$-$I^3_{d,R}$-${\rm PDF}$ &
 $G_F, M_W$ & $\alpha, M_Z, M_W, v_{q,L}, a_{q,L}$ \\
 \hline
    \end{tabular}
    \caption {\small \label{tab:fits}
               Summary of the main fit assumptions. In the fits, in addition
               to the free parameters listed in the first column, 
               the systematic
               correlation uncertainty parameters are allowed to 
               vary (see Table 2 in~\cite{h1ep320}).
               The fixed parameters are
               set to values taken from~\cite{pdg04} and $M_H$ is set to 
               $120\,{\rm GeV}$.
               }
\end{center}
\end{table*}

\section{Results}\label{sec:results}
\subsection{\boldmath Determination of Masses and $\sin^2\!\theta_W$}
\label{sec:mw}
\begin{figure}[htb] 
\begin{center}
\begin{picture}(50,125)
\put(-40,-10){\epsfig{file=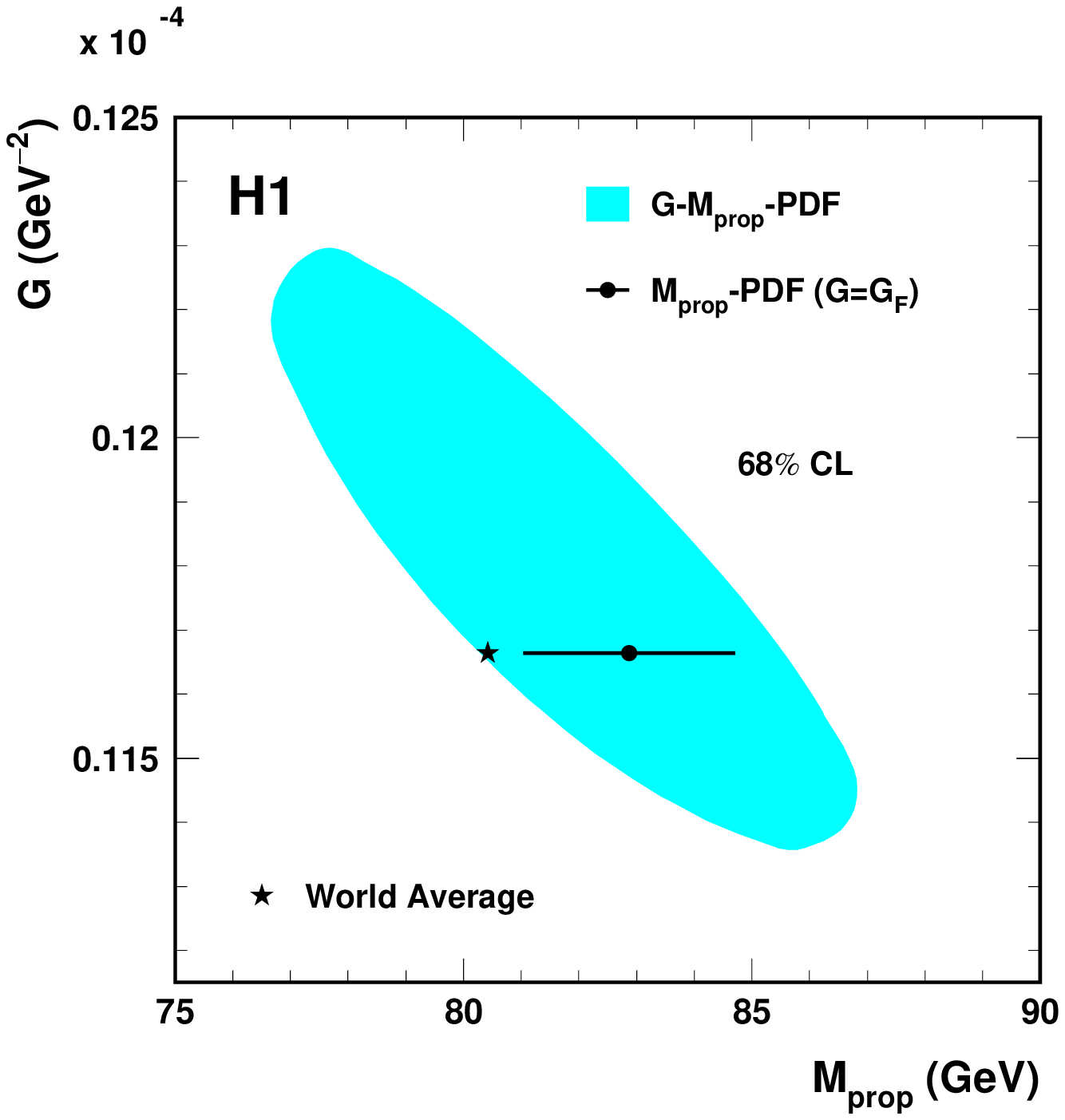,width=14cm}}
\end{picture}
\end{center}
  \caption{\label{fig:gf}
  The result of the fit to $G$ and $M_{\rm prop}$ at $68\%$ confidence 
  level (CL) shown as the shaded area. The world average values 
  are indicated 
  with the star symbol. Fixing $G$ to $G_F$, the fit 
  results in a measurement of the propagator mass $M_{\rm prop}$ shown
  as the circle with the horizontal error bars.
  }
\end{figure}

The cross section data allow a simultaneous determination of 
$G_F$ and $M_W$ and of the PDFs as independent parameters
(fit $G$-$M_{\rm prop}$-${\rm PDF}$ in Table~\ref{tab:fits}). 
In this fit, the parameters $G_F$ and $M_W$ in Eqn.(\ref{eqn:xscc}) 
are considered to be a normalisation variable $G$ and a propagator mass 
$M_{\rm prop}$, respectively, independent of the SM.
The sensitivity to $G$ according to Eqn.(\ref{eqn:xscc}) results from
the normalisation of the CC cross section whereas the sensitivity
to $M_{\rm prop}$ arises from the $Q^2$ dependence.
The fit is performed including 
the NC cross section data 
in order to constrain the PDFs. 
The result of the fit to $G$ and $M_{\rm prop}$ 
is shown in Fig.~\ref{fig:gf} as the shaded area.
The $\chi^2$ value per degree of freedom (dof) is $533.0/610=0.87$.
The correlation between $G$ and $M_{\rm prop}$ is $-0.85$, and is found to be
larger than the correlations with the QCD parameters~\cite{gfmw_cor}.
This determination of $G$ is consistent with the more precise value of 
$1.16637\cdot 10^{-5}\,{\rm GeV}^{-2}$ of $G_F$ obtained from 
the muon lifetime measurement~\cite{pdg04}, demonstrating the universality 
of the CC interaction over a large range of $Q^2$ values.

Fixing $G$ to $G_F$, one may fit the
CC propagator mass $M_{\rm prop}$ only.
For this fit ($M_{\rm prop}$-${\rm PDF}$), the EW parameters 
are defined in the MOMS scheme
and the propagator mass $M_{\rm prop}$ is considered to be 
independent of any other EW parameters.
Note that in the MOMS scheme, the use of $G_F$ makes the dependency of the CC 
and NC cross sections on $m_t$ and $M_H$ negligibly small.
The result of the fit, also shown 
in Fig.~\ref{fig:gf}, is
%
\begin{equation}
M_{\rm prop}=82.87\pm 1.82_{\rm exp}\left.^{+0.30}_{-0.16}\right|_{\rm model}\,{\rm GeV}\,.\label{eq:prop}
\end{equation}
Here the first error is experimental and the second corresponds to
uncertainties due to input parameters and model assumptions as introduced 
in Table~5 in~\cite{h1ep320}
(e.g. the variation of $\alpha_s=0.1185\pm 0.0020$). 
The $\chi^2$ value per dof is $533.3/611$.
If the PDFs are fixed in the fit, the experimental error on $M_{\rm prop}$ is reduced 
to $1.5\,{\rm GeV}$, which indicates that the correlation between $M_{\rm prop}$
and the QCD parameters is not very strong but not negligible either~\cite{bpthesis}. 
The determination given in
Eqn.(\ref{eq:prop}) represents the most accurate measurement 
so far of the CC propagator mass at HERA~\cite{h1ep300,h1nccc96,h1em320,
zeusccep300,zeusccem320,zeusccep320}.

The propagator mass $M_{\rm prop}$ measured here in the space-like region
can be compared with direct $W$ boson mass measurements obtained in 
the time-like region by the Tevatron and LEP experiments.
The value is consistent with 
the world average of 
$M_W=80.425\pm 0.038\,{\rm GeV}$~\cite{pdg04} within 1.3 standard deviations.

Within the SM, the CC and NC cross sections can be expressed in the OMS
scheme in which  
all EW parameters are determined by $\alpha$, $M_Z$ and $M_W$ together with
$m_t$ and $M_H$ in the loop corrections.
In this scheme, the CC cross section normalisation
depends on $M_W$ via the $G_F-M_W$ relation (Eqn.(\ref{eqn:gf})).
Some additional sensitivity to $M_W$ comes through the $M_W$ 
dependent terms (e.g., Eqn.(\ref{eqn:kappa})) in the NC cross section.
Fixing $m_t$ to its world average value of $178\,{\rm GeV}$~\cite{pdg04} 
and assuming $M_H=120\,{\rm GeV}$, the fit $M_W$-${\rm PDF}$ leads to
%
\begin{equation}
\label{eqn:mw}
M_W=80.786\pm 0.205_{\rm exp}\left.^{+0.048}_{-0.029}\right|_{\rm model}\pm0.025_{\delta m_t}-0.084_{\delta M_H}\pm 0.033_{\delta(\Delta r)}\,{\rm GeV}\,.
\end{equation}
Here, in addition to the experimental and model uncertainties, three other
error sources are considered: the uncertainty on the top quark
mass $\delta m_t=4.3\,{\rm GeV}$~\cite{pdg04}, 
a variation of the Higgs mass from $120\,{\rm GeV}$
to $300\,{\rm GeV}$ and the uncertainty of higher-order terms in 
$\Delta r$~\cite{bpthesis,kniehl96}. 
It should be pointed out that the result Eqn.(\ref{eqn:mw}) on $M_W$
is not a direct measurement 
but an indirect SM parameter determination which provides a consistency check
of the model.

Together with the world average value of 
$M_Z=91.1876\pm 0.0021\,{\rm GeV}$~\cite{pdg04}, 
the result obtained on $M_W$ from Eqn.(\ref{eqn:mw}) represents
an indirect determination of $\sin^2\!\theta_W$
in the OMS scheme (Eqn.(\ref{eqn:s2theta}))
%
\begin{equation}
\sin^2\!\theta_W=0.2151
\pm 0.0040_{\rm exp}\left.^{+0.0019}_{-0.0011}\right|_{\rm th}
\end{equation}
where the first error is experimental and the second is theoretical covering
all remaining uncertainties in Eqn.(\ref{eqn:mw}). The uncertainty due to
$\delta M_Z$ is negligible. 

Fixing $M_W$ to the world average value and assuming $M_H=120\,{\rm GeV}$,
the fit $m_t$-${\rm PDF}$ gives
$m_t=108\pm 44\,{\rm GeV}$
where the uncertainty is experimental.
The result represents the first determination of the top quark mass through
loop effects in the $ep$ data at HERA.

\subsection{\boldmath Determination of $v_{u,d}$ and $a_{u,d}$}
\label{sec:coupling}

At HERA, the NC interactions at high $Q^2$ receive contributions from 
$\gamma Z$ interference and $Z^0$ exchange (Eqns.(\ref{eq:f2},\ref{eq:xf3})). 
Thus the NC data
can be used to extract the weak couplings of up- and down-type quarks
to the $Z^0$ boson.
At high $Q^2$ and high $x$, where the NC $e^\pm p$ cross sections are 
sensitive to these couplings, the up- and down-type quark distributions 
are dominated by the light $u$ and $d$ quarks.
Therefore, this measurement can be considered to determine the light quark
couplings.
The CC cross section data help
disentangle the up and down quark distributions. 

\begin{figure}[htb] 
\begin{center}
\begin{picture}(50,160)
\put(-40,35){\epsfig{file=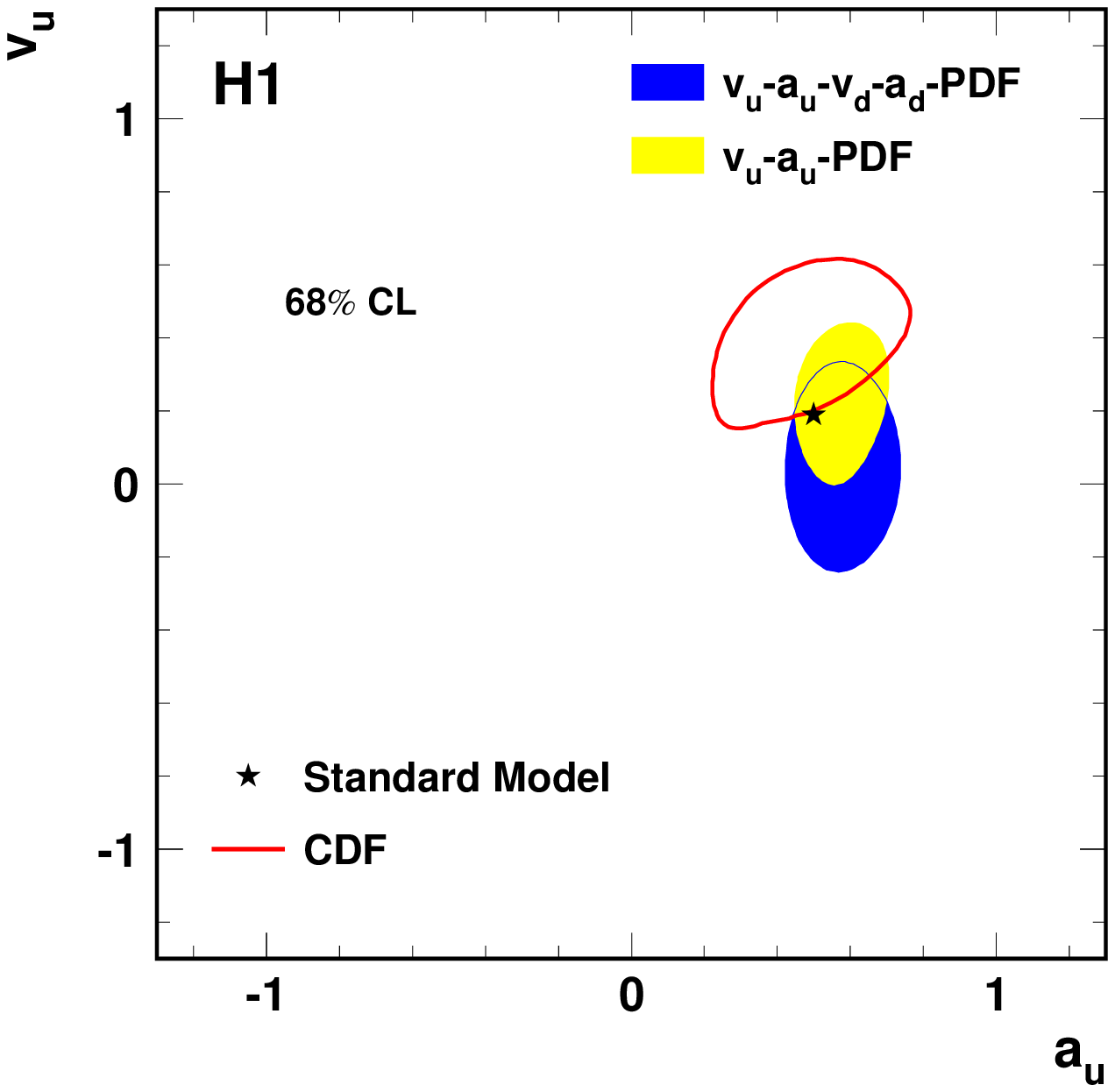,bbllx=0pt,bblly=0pt,bburx=594pt,bbury=842pt,width=15cm}}
\put(-40,-60){\epsfig{file=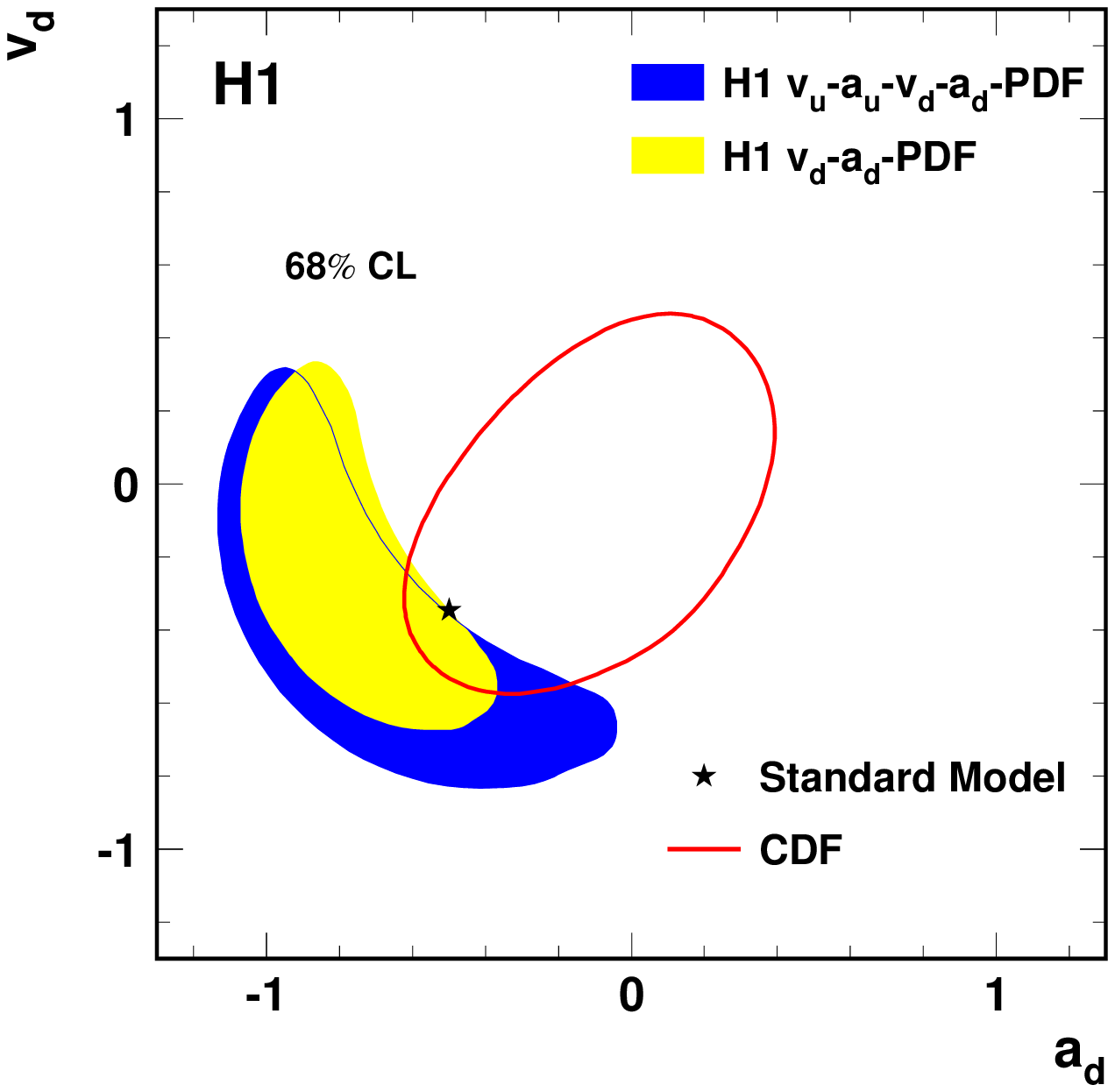,bbllx=0pt,bblly=0pt,bburx=594pt,bbury=842pt,width=15cm}}
\end{picture}
\end{center}
  \caption{\label{fig:coupling}
  Results at $68\%$ confidence level (CL) on the weak neutral current
  couplings of  $u$ (upper plot) and $d$ (lower plot) quarks to the $Z^0$ 
  boson determined in this analysis (shaded contours). 
  The dark-shaded contours correspond to results of a simultaneous fit of
  all four couplings and can be compared with those determined by 
  the CDF experiment (open contours). 
  The light-shaded contours correspond to 
  results of fits where either $d$ or $u$ quark couplings are fixed 
  to their SM values.
  The stars show the expected SM values.
  Preliminary contours (not shown) obtained from $e^+ e^-$ measurements at the $Z^0$ resonance
  can be found in~\cite{lepew}.
  }
\end{figure}

In this analysis (fit $v_u$-$a_u$-$v_d$-$a_d$-${\rm PDF}$), the vector and 
axial-vector dressed couplings of $u$ and $d$ quarks are treated as 
free parameters.
The results of the fit 
are shown 
in Fig.~\ref{fig:coupling}
and are given in Table~\ref{tab:coupling}.
The effect of the $u$ and $d$ correlation is illustrated in 
Fig.~\ref{fig:coupling}
by fixing either $u$ or $d$ quark couplings to their SM values 
(fits $v_d$-$a_d$-${\rm PDF}$ and $v_u$-$a_u$-${\rm PDF}$). 
The precision is better for the $u$ quark as expected.
The superior precision for $a_u$ comes from 
the $\gamma Z$ interference contribution $xF_3^{\gamma Z}$ 
(Eqn.(\ref{eq:xf3})). The $d$-quark couplings $v_{d}$ and $a_d$
are mainly constrained by the $Z^0$ exchange term $F_2^Z$ 
(Eqn.(\ref{eq:f2})). These differences in sensitivity result in the different
contour shapes shown in Fig.~\ref{fig:coupling}.


\begin{table*}[htb]
  \renewcommand{\doublerulesep}{0.4pt}
  \renewcommand{\arraystretch}{1.2}
 \vspace{-0.1cm}

\begin{center}
    \begin{tabular}{|c|c|c|c|c|c|}
      \hline
 Fit & $a_u$ & $v_u$ & $a_d$ & $v_d$ & $\chi^2/{\rm dof}$ \\ \hline
 $v_u$-$a_u$-$v_d$-$a_d$-${\rm PDF}$ & $0.56\pm 0.10$ & $0.05\pm 0.19$ &
  $-0.77\pm 0.37$ & $-0.50\pm 0.37$ & $531.7/608$ \\
 $v_u$-$a_u$-${\rm PDF}$ & $0.57\pm 0.08$ & $0.27\pm 0.13$ & $-$ & $-$ & 
  $534.1/610$ \\
 $v_d$-$a_d$-${\rm PDF}$ & $-$ & $-$ & $-0.80\pm 0.24$ & $-0.33\pm 0.33$ &
  $532.6/610$ \\\hline
 SM value & $0.5$ & $0.196$ & $-0.5$ & $-0.346$ & $-$ \\\hline
    \end{tabular}
    \caption {\small \label{tab:coupling}
               The results of the fits to the weak neutral current couplings 
               in comparison with their SM values. The correlation between the
               the fit parameters may be
               found in~\cite{cor-coupling}.
               }
\end{center}
\end{table*}

The results do not depend significantly on the low $x$ data,
nor on the assumptions on the parton distributions at low $x$ where
DGLAP may fail.
This was checked by performing two other fits, one for which the
data at $x \leq 0.0005$ are excluded, and another one for which
the normalisation constraint on the low $x$ behaviour of the anti-quark
distributions is relaxed\footnote{
 Further relations between the QCD parameters are given by
 sum rules and thus were not relaxed. The number of parameters which
 determine the parton densities was unchanged with respect to the QCD
 fit performed in~\cite{h1ep320}, where it was obtained using a
 well-defined $\chi^2$ minimisation procedure.
}.
This limited influence of the low $x$ region on the values of the
fitted EW couplings is partly due to the fact that electroweak effects
are most prominent at large $x$ and $Q^2$. Moreover the correlations
between the fitted couplings and the PDF parameters are moderate,
amounting to at most $21\%$~\cite{cor-coupling}.

The results from this analysis are also
compared in Fig.~\ref{fig:coupling} with 
similar results obtained recently by the CDF experiment~\cite{cdf}.
The HERA determination has comparable precision to that from the Tevatron.
These determinations are sensitive to $u$ and $d$ quarks separately, 
contrary to other measurements of the light quark-$Z^0$ couplings in 
$\nu N$ scattering~\cite{nuN} and atomic parity violation~\cite{apv} on
heavy nuclei. 
They also resolve any sign ambiguity and the ambiguities between $v_u$ and $a_u$ of the determinations
based on observables measured at the $Z^0$ resonance~\cite{lepew}.

\begin{figure}[htb] 
\begin{center}
\begin{picture}(50,100)
\put(-34,-10){\epsfig{file=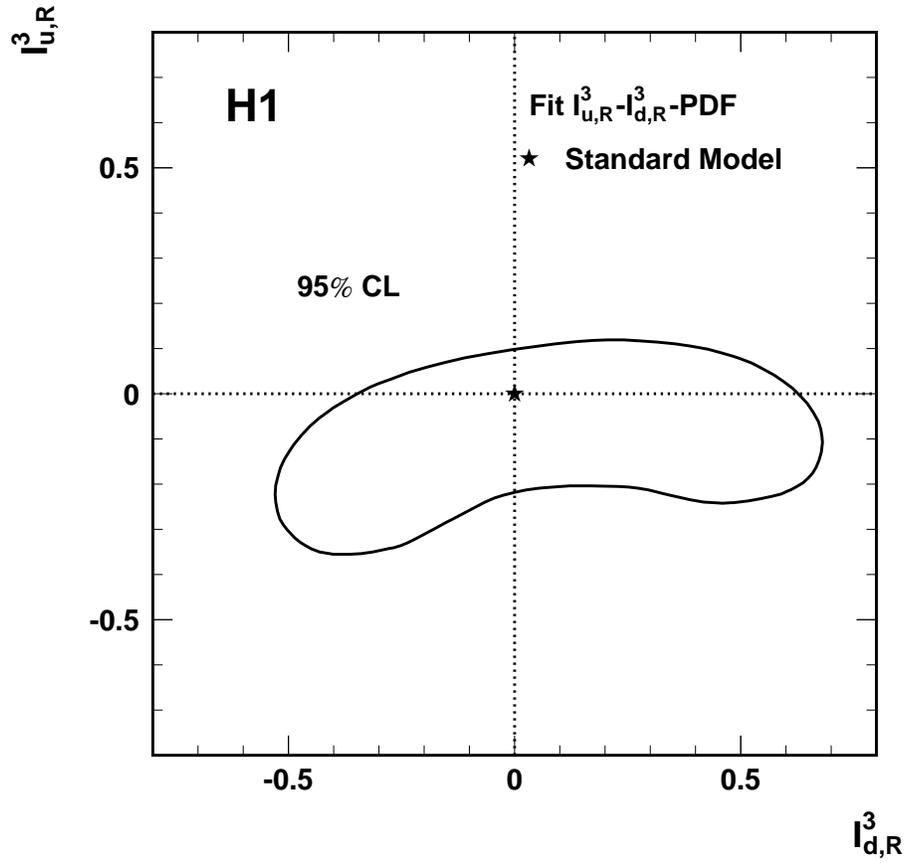,width=14cm}}
\end{picture}
\end{center}
  \caption{\label{fig:t3r}
  The result of the fit to the right-handed weak isospin charges $I^3_{u,R}$ 
  and $I^3_{d,R}$ at $95\%$ confidence level 
  (CL). In the SM the right-handed charges are zero (star symbol).}
\end{figure}

In more general EW models which consider other weak isospin multiplet structure,
the vector and axial-vector couplings in Eqns.(\ref{eqn:vf},\ref{eqn:af})
are modified in the following way~\cite{Klein:1979dp}
%
\begin{eqnarray}
v_q&=&I^3_{q,L}+I^3_{q,R}-2e_q\kappa_q\sin^2\!\theta_W\\
a_q&=&I^3_{q,L}-I^3_{q,R}\,.
\end{eqnarray}
Fixing $I^3_{q,L}$ and $\sin^2\!\theta_W$
to their SM values, a fit to $I^3_{u,R}$ and $I^3_{d,R}$ is performed
(fit $I^3_{u,R}$-$I^3_{d,R}$-${\rm PDF}$).
The results are shown in Fig.~\ref{fig:t3r}.
Both quantities are consistent with the SM prediction $I^3_{q,R} = 0$,
although 
the precision is not yet sufficient to exclude
a contribution of quarks in right-handed multiplets.

\section{Conclusion} \label{sec:conclusion}
Using the neutral and charged current cross section data recently published
by H1, combined electroweak and QCD fits have been performed.
In this analysis the correlation between the electroweak
and parton distribution parameters is taken into account and a set of
electroweak theory parameters is determined for the first time at HERA.

Exploiting the $Q^2$ dependence of the charged current data, 
the propagator mass has been measured with the result
$M_{\rm prop}=82.87\pm 1.82_{\rm exp}\left.^{+0.30}_{-0.16}\right|_{\rm model} {\rm GeV}$.
Within the Standard Model framework,
the $W$ mass has been determined to be
$M_W=80.786\pm0.205_{\rm exp}\left.^{+0.063}_{-0.098}\right|_{\rm th}\,{\rm GeV}$
in the on-mass-shell scheme.
This mass value has also been used to derive an indirect
determination of $\sin^2\!\theta_W$ yielding 
$0.2151\pm 0.0040_{\rm exp}\left.^{+0.0019}_{-0.0011}\right|_{\rm th}$.
Furthermore, a result on the top quark mass via electroweak 
effects in $ep$ data has been obtained.

The vector and axial-vector weak neutral current couplings of 
$u$ and $d$ quarks to the $Z^0$ boson have been determined at HERA for the
first time.
A possible contribution to the weak neutral current couplings
from right-handed current couplings has also been studied.
All results are consistent with the electroweak Standard Model.

\section*{Acknowledgements}
We are grateful to the HERA machine group whose outstanding
efforts have made this experiment possible.
We thank the engineers and technicians for their work in constructing 
and maintaining the H1 detector, our funding agencies for
financial support, the DESY technical staff for continual assistance
and the DESY directorate for support and for the
hospitality which they extend to the non DESY
members of the collaboration. It is our pleasure to thank H.~Spiesberger
for helpful discussions.

\newpage

%

%

%

\end{document}